# MAIN PARAMETERS OF THE LINAC-RING TYPE CHARM-TAU FACTORY

A. K Çiftçi, E. Recepoğlu, Physics Dept., Science Faculty, Ankara University, 06100, Tandogan, Ankara, TURKEY

Ö. Yavaş, Dept. of Eng. Physics, Faculty of Engineering, Ankara University, 06100, Tandogan, Ankara, TURKEY

S. Sultansoy, Physics Dept., Faculty of Arts and Sciences, Gazi University, 06500, Teknikokullar, Ankara, TURKEY

*Abstract*

Main parameters of the linac-ring type charm-tau factory are discussed. Different sets of parameters for an electron linac and a positron ring have been considered. It is shown that $L = 10^{33}$ cm$^{-2}$s$^{-1}$ and even more can be achieved. The physics goals of this machine in investigation for charmed particles and tau lepton properties are briefly discussed. Advantages of the proposed machine in comparison with standard (ring-ring) type charm-tau factory proposals will be the possibility of the search for rare decays of tau lepton and charm quark and study of $D^0 - \overline{D}^0$ oscillations.

## 1 INTRODUCTION

An old idea of colliding of the electron beam from linac with a beam stored in a ring [1] is widely discussed during the last decade with two purposes:

- to achive the TeV energy scale in lepton-hadron and photon-hadron collisions (see review articles [2] and references therein),
- to construct high luminosity particle factories, namely, B-factory [3], φ-factory, [4,5] c–τ-factory [6] etc.

Concerning the first direction, TESLA⊗HERA based ep, γp, eA and γA colliders are included in TESLA project [7]. And, Linac⊗LHC based ep, γp, eA, FELγA colliders [8] can be considered as the next step. On the other hand, linac-ring type B-factory has lost its attractiveness with KEK-B [9] and PEP-B [10] colliders coming into operation.

In this paper, we show that linac-ring type particle factories still are the matter of interest considering charm and tau options. In section 2, we present general investigation of beam dynamics aspects of linac-ring type colliders. Proposed parameters of linac-ring type charm and tau factories are discussed in section 3.1 and 3.2, respectively. In the final section, we give some concluding remarks.

## 2 GENERAL CONSIDERATIONS

From the point of view of particle physics, there are two most important collider parameters: center of mass energy and luminosity. For ultra-relativistic colliding beams, center of mass energy is given by

$$\sqrt{s} = 2\sqrt{E_1 E_2} \ . \qquad (1)$$

In our case, $E_1$ is the energy of electrons accelerated in the linac and $E_2$ is the energy of positrons stored in the ring. For charm factories, it is important to have $\Delta(\sqrt{s}) < \Gamma$ in order to use the advantage of resonant production of $\psi(3S)$ mesons: $m_{\psi(3S)} = 3769.9 \pm 2.5$ MeV with $\Gamma_{\psi(3S)} = 23.6 \pm 2.7$ MeV [11]. This condition is not so crucial for tau factory because tau leptons are produced in pairs.

The luminosity of $e^-e^+$ collisions is given by

$$L = \frac{N_e N_p}{2\pi \sqrt{(\sigma_{xe}^2 + \sigma_{xp}^2)(\sigma_{ye}^2 + \sigma_{yp}^2)}} f_c H_D \qquad (2)$$

where $N_e$ is number of electrons per bunch, $N_p$ is number of positrons per bunch, $\sigma_{x,y}$ are horizontal and vertical beam sizes, $f_c$ is collision frequency. $H_D$ is luminosity enhancement factor which is calculated by using GUINEA-PIG beam-beam simulation program [12].

The first restrictive limitation for electron beam is beam power

$$P_e = N_e E_e f_c \qquad (3)$$

which determines the maximum value of $N_e f_c$ in Eq. (2). We have chosen $P_e \leq 10$ MW.

The maximum number of electrons per bunch is determined by the beam-beam tune shift limit of the positron beam

$$\Delta Q_p = \frac{N_e r_0 \beta_p^*}{2\pi \gamma_p \sigma_{ye}(\sigma_{xe} + \sigma_{ye})} \quad (4)$$

where $r_0 = 2.81 \times 10^{10}$ m is the clasical radius of the electron, $\gamma_p$ is the Lorentz factor of the positron beam and $\beta_p^*$ is the beta function at collision point. Generally accepted beam-beam tune shift value for positrons in case of ring-ring colliders is $\Delta Q_p \leq 0.06$. This limit value can be a little bit larger for linac-ring type colliders.

Parameters of positron beams are constracted by the disruption D of electrons, which is defined as the ratio of the positron bunch length to the electron focal length

$$D_{ye} = \frac{\sigma_{zp}}{f_{ye}} = \frac{2r_0 N_p \sigma_{zp}}{\gamma_e \sigma_{yp}(\sigma_{xp} + \sigma_{yp})} \quad (5)$$

where $\gamma_e$ is the Lorentz factor of the electron beam and $\sigma_{zp}$ is the positron bunch length. The analysis performed for linear colliders shows $D_{xe}=D_{ye}=D=25$ is acceptable to avoid kink instability [13]. In this study we consider the round beam case: $\sigma_{xe} = \sigma_{ye} = \sigma_e$ and $\sigma_{xp} = \sigma_{yp} = \sigma_p$.

Another restrictive condition which limits number of particles in a positron bunch is microwave instability. The microwave instability for $|Z/n| \sim 1\,\Omega$ consistent with SSRF, limits the number of positrons per bunch to $\sim 2.5 \cdot 10^{10}$ for charm factory and $2.1 \cdot 10^{10}$ for tau factory. The transverse mode coupling instability limits of the positrons bunch population is higher than the microwave instability limit. Therefore it can be ignored. Scaled from SSRF proposal bunch lengthening due to potential well distortion is less than 10%. As a result, number of positrons per bunch in our parameter sets are below of limits of these instabilities.

## 3 PARAMETER SETS

### 3.1 Linac-Ring Type Charm Factory

Recently CLEO-c (see [14] and references therein) proposal has been approved in order to explore the charm sector starting early 2003. With a necessary upgrade, expected machine performance for CLEO-c will be $3 \times 10^{32}$ cm$^{-2}$s$^{-1}$ at $\sqrt{s} = 3.77$ GeV. In Table I, we present proposed parameters for linac-ring type charm factory option. As one can see $10^{33}$ cm$^{-2}$s$^{-1}$ can be achieved which exceeds the CLEO-c design luminosity by more than factor three.

Table 1: Main Parameters of Charm and Tau Factory

|  | Charm | Tau |
|---|---|---|
| $E_e$ (GeV) | 1.42 | 2.1 |
| $E_p$ (GeV) | 2.5 | 2.1 |
| $N_e$ ($10^{10}$) | 0.1 | 0.07 |
| $N_p$ ($10^{10}$) | 2 | 2 |
| $\beta_e / \beta_p$ (cm) | 0.45/0.45 | 0.45/0.45 |
| $\varepsilon_e^N / \varepsilon_p^N$ ($\mu$m rad) | 2.15/4 | 2.65/2.25 |
| $\sigma_{ze} / \sigma_{zp}$ (cm) | 0.1/0.45 | 0.1/0.45 |
| $\sigma_e / \sigma_p$ ($\mu$m) | 1.86/1.91 | 1.70/1.56 |
| $\Delta Q_p$ ($10^{-3}$) | 0.059 | 0.059 |
| Disruption | 24.8 | 25 |
| $f_c$ (MHz) | 30 | 30 |
| $H_D L$ ($10^{33}$ cm$^{-2}$ s$^{-1}$) | 1 | 1 |
| LinacBeam Power (MW) | 6.8 | 7 |

In the case of charm factory, it is important to obey condition $\Delta(\sqrt{s}) < \Gamma$. The expected luminosity spectrum $dL/dW_{cm}$ is plotted in Figure 1. We have used GUINEA-PIG simulation program [12] with $\Delta E_{e^+}/E_{e^+} = \Delta E_{e^-}/E_{e^-} = 10^{-3}$. It is seen that center of mass energy spread is well below $\Gamma_{\psi(3S)} \approx 24$ MeV. Therefore, we can use the well known Breit-Wigner formula

$$\sigma_{BW} = \frac{12\pi}{m_{\psi(3S)}^2} B_{in} B_{out} \quad (6)$$

where $B_{in}$ and $B_{out}$ are the branching fractions of the resonance into the entrance and exit channels. By taking, $\text{Br}(\psi(3S) \to e^+ e^-) \approx 10^{-4}$.

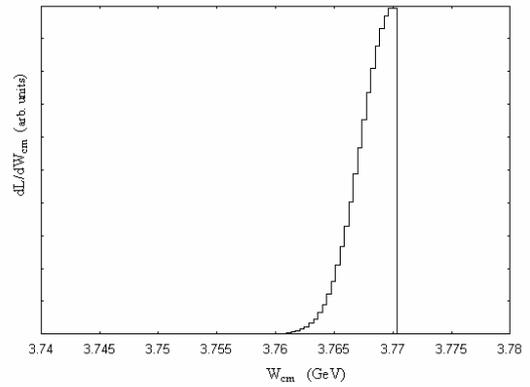

Figure 1. Luminosity Spectrum for Charm Factory

We obtain about $10^9$ expected number of $\psi(3S)$ per working year ($10^7$ s). Let us mentioned that $D\overline{D}$ decay mode is the dominant channel for $\psi(3S)$ decays. An additional advantage of the proposed charm factory is the asymmetric kinematics. This feature will be important in investigations of $D - \overline{D}^0$ oscillations and CP-violations in charmed particle decays.

### 3.2 Linac-Ring Type Tau Factory

The cross section of the process $e^+e^- \rightarrow \tau^+\tau^-$ for $s \ll m_Z^2$ is given by

$$\sigma = \frac{2\pi}{3}\frac{\alpha^2}{s}\beta(3-\beta^2) \approx \frac{43.4\text{nb}}{s(\text{GeV})^2}\beta(3-\beta^2) \quad (7)$$

where $\beta = \sqrt{1-4m_\tau^2/s}$ and $\alpha$ is the fine structure constant. The maximum value of $\sigma = 3.56$ nb is achieved at $\sqrt{s} \approx 4.2$ GeV. In difference from charm factory, in the case of $\tau$-factory we have consider the symmetric option $E_{e^-} = E_{e^+} = 2.1$ GeV). Proposed set of parameters is given in the last column.

One can see that linac-ring type $\tau$-factory will produce $\approx 4\cdot 10^7$ $\tau^+\tau^-$ pair per working year, which exceeds by one order the statistics obtained at LEP and CLEO up to now.

Until this point, we have assumed that lower limit on $\beta^*$ in both parameter sets is given by the positron beam bunch length (Hourglass effect). This limitation can be relaxed by applying a "dynamic" focusing scheme, where the positron beam waist travels with the e- bunch during collision [15]. This scheme requires a pair of pulsed RF-quadropoles to be installed on both side of interaction region. In this way, the limitation on $\beta^*$ becomes equal to electron bunch length. This corresponds to upgrade of the luminosities on Table I by factor 4.5.

## 4. CONCLUSION

We have shown that linac-ring type machines will give an opportunity to achieve $L = 10^{33}$ cm$^{-2}$s$^{-1}$, which essentially exceeds the luminosity values of existing and proposed standard (ring-ring type) charm and tau factories. This leads to an obvious advantage in search for rare decays. Another important feature of linac-ring type charm factory is the asymmetric kinematics. This will be important in investigation of oscillations and CP-violation in strange and charm sector of the SM.